\documentstyle[12pt]{article}
\title{Note on reversibility of quantum jumps\footnote{Published in 
Physics Letters A 222, No. 3, pp. 137-140 (1996)}}
\author{Michael B. Mensky\thanks{E-mail: mensky@sci.lebedev.ru} \\
\normalsize \it P.N.Lebedev Physical Institute, 117924 Moscow, Russia}
\date{}

\begin{document}
\maketitle
\abstract{It has been recently proved that a quantum jump may be 
reversed by a unitary process provided the initial state is restricted 
by some conditions. The application of such processes for preventing 
decoherence, for example in quantum computers, was suggested. We shall 
show that in the situation when the quantum jump is reversible it 
supplies no information about the initial state additional to the 
information known beforehand. Therefore the reversibility of this type 
does not contradict the general statement of quantum measurement 
theory: a measurement cannot be reversed. As a consequence of this, 
the coherence of a state (say, in a quantum computer) cannot be 
restored after it is destroyed by dissipative processes having a 
character of measurement.}

\vspace{0.5cm}
\vspace{0.5cm}

The problem of preventing quantum decoherence in real systems became 
recently important in connection with the question about realizability 
of quantum computers \cite{qu-comp,Landauer}. Decoherence as a 
physical phenomenon may be considered in a more general framework of 
theory of quantum noise \cite{qu-noise}. However, decoherence as a 
physical process arising in the course of a quantum measurement, has 
interesting specific aspects. Some of them will be discussed here, 
with important conclusions about possibility to prevent decoherence.    

Mabuchi and Zoller has considered recently \cite{Zoller} a specific 
type of dissipation processes that may be characterized as a quantum 
jump i.e. disappearing of a photon, for example its absorption by a 
detector. The quantum jump may be described by an annihilation 
operator $c$ (from the pair $c, c^{\dagger}$) as the transition 
$|\psi\rangle\to c|\psi\rangle$. It has been proved in  \cite{Zoller} 
that the quantum jump can be reverted with the help of an unitary 
evolution provided the system has been before the jump in a state from 
a certain subspace. As a result, the initial state may be restored 
coherently, with the same phase relations as before the jump. This 
process was suggested as a possible mechanism for preventing 
decoherence, with possible application in theory of quantum computers. 

A quantum jump is an example of a dissipative process. Therefore,  the 
result of Mabuchi and Zoller \cite{Zoller} proves that some of 
dissipative processes may be reverted. Then the procedure providing 
the inversion of quantum jumps may serve as a method of preventing 
dissipation. We shall show however that the dissipation prevented in 
this way is not accompanied by obtaining new information and therefore 
cannot be identified with the decoherence arising in the process of a 
quantum measurement. It seems plausible that this is a general 
situation: decoherence cannot be inversed if any information is 
supplied by the process leading to this decoherence. This essentially 
restricts applicability of the procedure of Mabuchi and Zoller.   

Our goal is therefore to show that the inversion of a quantum jump is 
possible only in the case when the jump supplies no information 
(additional to the information we had already before the jump), 
therefore it cannot be considered to be a measurement. The reversible 
dissipation is not a decoherence arising in the course of a quantum 
measurement.  

1. Let the quantum jump be described by the annihilation operator $c$
(from the pair of creation-annihilation operators $c^{\dagger}, c$). As
it has been proved in \cite{Zoller}, the quantum jump $c$ may be reverted
(i.e. the initial state of the system recovered) with the help of a
unitary evolution, provided that the initial state belongs to some 
subspace of the state space ${\cal H}$.

This means that the action of the operator $c$ on an arbitrary state
from the specified subspace is identical with the action of some
unitary operator $U$.  Recovering of the initial state is then
possible with the help of the evolution described by the operator
$U^{\dagger}=U^{-1}$. 

For the goal of the general theoretical analysis of this situation, we 
shall denote by ${\cal H}_1$ the subset of all vectors with this 
property, so that  
\begin{equation}
\left. c\right|_{{\cal H}_1 } =\left. U\right|_{{\cal H}_1 }.
\label{equiv}\end{equation}

If the quantum jump $c$ could give some information about the initial 
state, then it might be interpreted as a measurement. The 
contradiction with quantum measurement theory could arise in this 
case: the effect of the measurement on the system might be completely 
discharged by a certain unitary evolution. Our task is to show that 
this is not the case. We shall prove that, as a consequence of 
Eq.~(\ref{equiv}), the event of the quantum jump $c$ gives no 
information about the initial state (other than the information 
following from the fact that the initial state belongs to ${\cal 
H}_1$). Therefore this event cannot be interpreted as a measurement.

For this end, we shall derive some properties of the states belonging to ${\cal H}_1$ and prove that the quantum jump gives no information about the initial state besides that this state had these properties. 

First of all, it follows directly from Eq.~(\ref{equiv}) that for an 
arbitrary  vector $|\psi\rangle\in{\cal H}_1$ the following equations 
are satisfied: 
\begin{equation}
c|\psi\rangle =U|\psi\rangle, \quad \langle\psi | c^{\dagger} 
= \langle\psi | U^{\dagger}. 
\label{num}\end{equation}
This means (because of unitarity of $U$) that the mean photon number 
for an arbitrary state from the specified subset, 
$|\psi\rangle\in{\cal H}_1$, is equal to unity:
\begin{equation}
\langle\psi|N|\psi\rangle=1
\label{mean-num}\end{equation}
where $N=c^{\dagger}c$ is an operator of the photon number. 

Let us expand the state $|\psi\rangle$ in a series of terms
corresponding to definite photon numbers:
\begin{equation}
|\psi\rangle
=c_0|\psi_0\rangle + c_1|\psi_1\rangle + c_2|\psi_2\rangle + \dots
  + c_n|\psi_n\rangle + \dots
\label{num-expand}\end{equation}
where $|\psi_n\rangle$ is a (normalized) state with $n$ photons. Then
Eq.~(\ref{mean-num}) reads as follows:
\begin{equation}
p_1 + 2 p_2 + 3 p_3 + \dots + n p_n + \dots =1
\label{mean-expand}\end{equation}
with positive numbers $p_n=|c_n|^2$. For this equation being 
fulfilled, at least one of the numbers $p_1$, $p_2$, \dots $p_n$, 
\dots must be non-zero. Therefore, the state $|\psi\rangle\in{\cal 
H}_1$ cannot be vacuum. The expansion (\ref{num-expand}) must contain 
at least one non-vacuum component.

The quantum jump $c$ diminishes the number of photons by unity. 
Therefore, the fact that the jump (the click of the detector) 
occurred, gives the information that the initial state has contained 
not less than one photon (could not be vacuum). However we know this 
already from Eq.~(\ref{mean-expand}). The event of the jump gives no 
new information and cannot be considered to be a measurement (provided 
we know already that the system has been in the subset ${\cal H}_1$ 
before the jump).

Of course, if two quantum jumps occur, this will supply some new 
information: that the number of photons was not less than 2. This 
could be a measurement. This however again leads to no contradiction, 
because the action of the operator $c^2$ (describing a double jump) is 
not equivalent to the action of a unitary operator even in the subset 
${\cal H}_1$. The double jump is a measurement, and its effect cannot 
be discharged by a unitary evolution. It is irreversible, in complete 
correspondence with general principles of quantum theory of 
measurements.

2. Let us suppose now that not only a single jump, but also a double jump
may be unitarily reversed provided the system has been in the subset
${\cal H}_2$ before the jumps. This means that two unitary operators
$U_1, U_2$ exist such that
\begin{equation}
\left.c\right|_{{\cal H}_2 } =\left. U_1\right|_{{\cal H}_2 }, 
\quad  
\left.  c^2\right|_{{\cal H}_2 }=\left. U_2\right|_{{\cal H}_2 }.
\label{2-equiv}\end{equation}
Then the following relations may be readily derived for an arbitrary 
state from the subset,  $|\psi\rangle\in{\cal H}_2$:
\begin{equation}
\langle\psi|N|\psi\rangle=1, \quad \langle\psi|N(N-1)|\psi\rangle=1.
\label{2-mean-num}\end{equation}
Using the expansion (\ref{num-expand}), we  may rewrite the same
in the form
\begin{eqnarray}
p_1 + 2 p_2 + 3 p_3 + \dots + n p_n + \dots &=&1\nonumber\\
2 p_2 + 6 p_3 + \dots + n(n-1) p_n + \dots &=&1.
\label{2-mean-expand}\end{eqnarray}

It is seen from Eqs.~(\ref{2-mean-expand}), that there is at least 2 
photons in an arbitrary state of the subset ${\cal H}_2$ (i.e. such a 
state cannot be a superposition of the vacuum and the 1-photon state). 
Therefore, neither a single, nor a double jump give no additional 
information in the case when the effects of both a single jump and a 
double jump can be unitarily discharged. the single and double jumps 
are not in this case measurements. 

It is evident that the same consideration is applicable also to the
case of a multiple jump with an arbitrary multiplicity.

3. Quantum jump that means for example a click of a detector is 
considered usually as a sort of measurement in the sense that it 
supplies a new information. It has been shown above that the quantum 
jump may give no information if something is known about the initial 
state. In this case a measurement supplying nontrivial information 
might take place in the preceding step when the initial state had been 
prepared. This step should contain projection from the complete space 
of states ${\cal H}$ onto some subspace belonging to ${\cal H_1}$ (or 
${\cal H_2}$ if the situation with double jumps is considered) .  
After this preliminary measurement the quantum jump (or double jump) 
gives no new information. One can formally say that such a jump is a 
measurement, but it should be clearly understood that this measurement 
gives no additional information. 

This  is not at all astonishing and is in fact common in quantum 
theory of measurement. Indeed, the text-book example of a quantum 
measurement is the measurement of an observable $A$ with a discrete 
spectrum. If we have a series of repeated measurements of this type 
(beginning from an unspecified state), then only the first measurement 
supplies a non-zero information giving the measurement output $a_i$. 
All subsequent measurements of $A$ will give with certainty the same 
result. The situation is quite analogous to the above discussion of 
quantum jumps if the stage of the preparation of an initial state is 
taken into account.  

4. The above arguments support the general statement about
irreversibility of quantum measurements (in the case when the 
initial state is not specified). This is important in the
context of quantum computers and other devices depending on 
coherent character of their evolution.

One may hope to prevent decoherence resulting from dissipative 
processes, applying some or another correcting procedures, for example 
those proposed in \cite{Zoller}. It is shown in \cite{Zoller} that the 
initial state may be coherently restored after a certain dissipative 
processes. However the arguments of the present paper (apparently 
applicable to a more general situation) demonstrate that the 
restoration of the coherence is not always possible.

The class of dissipative processes leading to the irreversible 
decoherence may be specified by the concept of measurement or 
information. The restoration of the coherence turns out to be 
impossible if the dissipation is accompanied by obtaining information 
about the state of a quantum system.\footnote{Of course, we do not 
necessarily mean a measurement arranged on purpose. Instead, it may be 
an interaction with the environment (reservoir) that results in 
recording the information in the state of the environment even if 
nobody is interested in this information.} This essentially restricts 
the circle of situations in which recoherence is in principle 
feasible. This resulting restriction should be taken into account 
together with other principal difficulties in creating quantum 
computers \cite{Landauer}. 

\vspace{0.5cm}
\centerline{\bf ACKNOWLEDGEMENT}

The author is indebted to I.Bialynicki-Birula and
P.Zoller for stimulating discussions and W.Schleich for his kind
hospitality in Ulm University during the workshop of the Ulm and
Innsbruck quantum-optics groups.

\end{document}